\documentstyle[preprint,prb,aps]{revtex}
\long\def\title#1{{\Large\begin{center}#1\end{center}\par}}
\long\def\address#1{\begin{center}#1\end{center}\par}
\long\def\author#1{\begin{center}#1\end{center}\par}
\def\pacs{}
\tightenlines
\begin{document}

\draft

\title{Lipids diffusion mechanism of stress relaxation in a
 bilayer fluid membrane under pressure.}

\author{Sergei I. Mukhin and Svetlana V. Baoukina}
\address{Moscow Institute for Steel and Alloys,
Theoretical Physics Department, Leninskii pr. 4, 117936 Moscow,
Russia}

\vspace{3mm}
%\date{today}
\maketitle

\begin{abstract}
A theory of a lateral stress relaxation in a fluid bilayer
membrane under a step-like pressure pulse is proposed. It is shown
theoretically that transfer of lipid molecules into a strained
region may lead to a substantial decrease of the membrane's free
energy due to local relaxation of the stress. Simultaneously, this
same effect also causes appearance of the spontaneous curvature of
the membrane. Proposed stress relaxation mechanism may explain
\cite{6} recent experimental observations of a collapse of the
ionic conductance through the protein mechanosensitive channels
piercing the E-coli's membrane \cite{4,5}. The conductance
decreases within seconds after opening of the channels by applied
external pressure. The time necessary for a propagation of a
stress over the largest membrane's dimension is $5\div 6$ orders
of magnitude shorter. On the other hand, theoretical estimates of
the phospholipids diffusion time inside the strained area is
favorably comparable with the experimental data \cite{5}. Our
theory also predicts that the membrane's curvature increases
simultaneously with the stress relaxation caused by the lipids
diffusion. This effect is proposed for experimental check up of
the theoretical predictions.
\end{abstract}

\section{Introduction}
 The functioning of the protein channels (pores) in the cell membranes
 regulates flows of ions in- and out of the cell, thus
 influencing signal transmission processes in the neural networks
 of the biological systems \cite{1}. Therefore, the gating mechanisms
 of the
 channels are of great interest for the fundamental and practical
 matters, e.g. like enhancement of the efficiency of the general
 anesthesia methods \cite{2}. Mechanical stresses in cell membranes
 may cause opening of the protein mechanosensitive (MS) channels \cite{3},
 which hence play a role of mechanoreceptors in the cellular organisms.
 One of the biological objects convenient for experimental study
 is the large conductance MS channel (MscL) in the inner membrane
 of the Escherichia coli (E-coli) bacteria. It is possible to measure
 a single channel conductance, an find its dependence on
 the internal lateral tension in the membrane. The tension, which causes
 opening
 of the channel, could
 be evaluated using video microscopy measurements of the membrane curvature
 formed under the applied external pressure \cite{4}.

 The time-dependence of the membrane's conductance observed by C.C.
 Hase {\it{et al.}} \cite{5} reveals gradual collapse of the ionic
 currents through the MscL's within seconds after their opening.
 According to existing hypothesis \cite{6}, this may happen due to
 mutual slide of the lipid monolayers, constituting a
 bilayer membrane of the E-coli. This slide would then encourage
 a relaxation of the lateral stress inside the membrane, which is
 necessary for
 an opening of the MscL's. The slide of the monolayers should be
 encouraged by the same pressure pulse, which causes initial opening
 of the ionic
 channels.

 In this paper we propose a theory of such a stress-relaxation
 process and demonstrate that the slide of the monolayers under a constant
 external pressure is indeed thermodynamically advantageous. In order to
 show this, we calculate the free energy difference between two
 possible equilibrium states of a membrane under a pressure gradient pulse.

 The first ("intermediate") equilibrium state, is achievable shortly after
 the
 beginning of a pressure pulse $P$, within the time of propagation of the
 mechanical stress along a membrane $\Delta t > 10^{-6}$sec (evaluated
 below). In this state
 the initial numbers of the phospholipid molecules in the monolayers
 constituting
 the bilayer membrane, $N_1$ and $N_2$,
 yet, remain unchanged: $N_1=N_2=N_0$. Here $N_0$ is the
 number of the molecules in a monolayer before external pressure gradient
 is applied. We neglect a small probability \cite{safr,7} of the
 flip-flop process, i.e. exchange of lipids between the monolayers.
 Then, the tensions $T_1$, $T_2$, in both monolayers approximately
 coincide, since the radius $R$,
 of the membrane is of order $1\div 2 \mu m$, while its thickness $l_0$,
 is about $2 nm$
 \cite{5}.

 In the second, (final) equilibrium state,
 achievable within diffusion time $\tau\sim 1\mbox{sec} \gg \Delta t$,
 the extra lipid
 molecules from the outer patch of E-coli's material have enough time to
 be "sucked in"
 inside the concave-side monolayer, see Fig. 1, until the tension in the
 monolayer is relaxed at a finite curvature: $T_1 =0$. Then, the tension
 $T_2$ in the convex-side monolayer (which is stuck to the pipette and
 thus can not
 absorb extra
 molecules \cite{4,5}, see Fig. 1) alone maintains a mechanical
 equilibrium under the same external pressure difference $P$ across the
 membrane.
 In accord with the Laplace's law we have:

\begin{equation}
 T_1+T_2=T_2=Pr/2;\;\;\;\mbox{at}\;\; T_1=0.
 \label{laplace}
 \end{equation}

 \noindent Here $r$ is the radius of curvature. Next, we
 calculate self-consistently the free energy of the whole system:
 membrane+patch, in both equilibrium states.
 Important, the free energy in the relaxed, second state, proves
 to be lower
 than in the intermediate state with constrained numbers of lipids,
 Fig. 2.
 Choosing the second, relaxed
 state we then evaluate the number $\Delta N_1$ of the "sucked in"
 lipids and
 find, see Fig. 4, that it is of a macroscopic magnitude, e.g.
 $\Delta N_1\sim 15 {\%} N_1$. Based on these results and using
 experimentally determined diffusion coefficient of individual
 lipid molecules in bio-membranes at room temperature
 \cite{schmidt}
 $D\sim 1\div 20 \mu m^2/sec$, we estimate the characteristic
 diffusion time $\tau$:
\begin{equation}
\tau \sim \displaystyle\frac{\Delta N_1}{N_1}\frac{R^2}{D} \sim 1
\mbox{sec},
 \label{diffusion}
 \end{equation}

\noindent which is comparable by the order of magnitude with the
experimental data \cite{5}. Thus, we predict a macroscopic
increase of the number of lipid molecules $\Delta N_1$ in the
concave monolayer and, hence, propose to measure the concomitant
change of the weight of the part of the membrane sucked in in the
pipette during the decrease of the ionic conductance (in the setup
of the works \cite{4,5}).

Besides, as it is apparent from Fig. 2, the equilibrium curvature
of the membrane (indicated with the arrows) also changes
significantly as the membrane passes from the intermediate to the
final equilibrium state. Hence, it should be possible to check our
theory also by measuring the time-dependent increase of the
curvature, which should then accompany the decay of the ionic
conductance observed \cite{5} under a constant external pressure
gradient.

The plan of the article is as follows. In Section II we describe a
free energy functional in our model of a bilayer membrane and
derive the self-consistent equations for the membrane in the
equilibrium state under a constant pressure gradient. Both
versions of the equilibrium state discussed above are considered.
In Section III we give analytical and numerical solutions of the
derived equations and discuss in detail corresponding calculated
dependencies. Finally, we discuss in the last Section IV possible
experimental verifications and future improvements of the theory.

\section{The model free energy}

We use the microscopic model of a bilayer membrane as described
e.g. in \cite{safr,7}. First write a single i-th monolayer
($i=1,2$) free energy per lipid molecule $f_i$:

\begin{eqnarray}
f_i=f_{si}+f_{hi}+f_{ti}\equiv \gamma a_i+\frac{C}{a_i}+
\displaystyle\frac{k_{s}}{2}(l_i-l_s)^2\label{fren}
\end{eqnarray}

\noindent Here $f_{si}$ is the external surface energy of the
monolayer, which increases with the area per lipid molecule $a_i$
due to the hydrocarbon tails interaction energy with the solvent.
The short-range repulsion between the polar molecular heads at the
surface of the membrane is represented by the second term in Eq.
(\ref{fren}). The third term, though looks like a spring elastic
energy (with not stretched spring length $l_s$), codes for the
entropic  repulsion between the hydrocarbon tails in the depth of
each monolayer \cite{7}. In what follows we omit the second term
in Eq.(\ref{fren}) on the empirical grounds \cite{7} relevant for
the chain molecules in phospholipid bilayers. There is no
interface energy term included in (\ref{fren}) for the internal
surfaces of the monolayers forming the bilayer, as the latter are
not reachable for the solvent molecules. We also assume a
vanishing interdigitation of the tails between the adjoint
monolayers. The length of a lipid molecule $l_i$, i.e. the
thickness of the i-th monolayer, is not an independent variable
from the aria per molecule $a_i$. The volume $v$ per lipid
molecule is conserved, i.e. the latter could be considered as
incompressible \cite{safr,7}. Due to this conservation condition
the length $l_i$ is related to the mean and gaussian curvatures
$H$ and $K$ at the interface between the monolayers by a well
known differential geometry formulas \cite{safr}:

\begin{eqnarray}
&& l_1=l_{01}-l_{01}^2H+\displaystyle\frac{2}{3} l_{01}^3K
\label{curvs1}\\
 && l_2=l_{02}+l_{02}^2H+\displaystyle\frac{2}{3}
l_{02}^3K \label{curvs2}
\end{eqnarray}

\noindent Here the change of sign in front of $H$-term in the
equation (\ref{curvs2}) relative to that in equation
(\ref{curvs1}) is due to a simple geometrical fact that the {\em
external} normal vector to e.g. the layer 1 is simultaneously an
{\em internal} normal vector to the layer 2 at their mutual
interface. Different $l_{0i}$ parameters just reflect the
incompressibility of the lipid molecules mentioned above:

\begin{eqnarray}
l_{0i}=\frac{v}{a_i} \label{lis}
\end{eqnarray}

\noindent Substituting  Eqs. (\ref{curvs1}),(\ref{curvs2}) into
the third term of Eq. (\ref{fren}) one finds contributions to the
free enrgy from the tails in the form:

\begin{eqnarray}
&&f_{t1}=\frac{k_sl_{01}^4}{2}\left[(H-C_{01})^2+
\frac{4}{3}C_{01}l_{01}K\right]
\label{ftis1}\\
&&f_{t2}=\frac{k_sl_{02}^4}{2}\left[(H+C_{02})^2+
\frac{4}{3}C_{02}l_{02}K\right]
\label{ftis2}
\end{eqnarray}

\noindent where parameters $C_{0i}$ have meaning of the local
spontaneous curvatures of the $i$-th layer:

\begin{eqnarray}
C_{0i}=\frac{l_{0i}-l_s}{l_{0i}^2} \label{cis}
\end{eqnarray}

We readily recognize the Helfrisch's  formula \cite{4b} for the
free energy of a membrane in Eqs. (\ref{ftis1}), (\ref{ftis2}).
In
a symmetric bilayer case $N_1=N_2$ (where again $N_1$ and $N_2$
signify the numbers of lipid molecules in the 1-st and 2-nd
monolayer respectively), with equivalent conditions on the
opposite surfaces of the membrane $a_1=a_2$, the linear in $H$
terms in Eqs. (\ref{ftis1}), (\ref{ftis2}) would cancel in the
total free energy $F$:

\begin{eqnarray}
F=\sum_{i=1,2}F_i=\sum_{i=1,2}N_if_i \label{frent}
\end{eqnarray}

\noindent Nevertheless, as shown below, a linear in $H$ term may
arise when the up {\it vs} down symmetry of a membrane is broken,
e.g. by applied pressure gradient in combination with the
different boundary conditions at the monolayers peripheries, see
Fig. 1.

To make the whole idea transparent we restrict our present
derivation to the case of a spherically "homogeneous
distributions" of molecules, i.e. considering $a_i$'s as being
different for the different indices i's, but position-independent
within i-th monolayer. Below, we neglect the inhomogeneous
distribution of the strain across the thickness of each monolayer.
Also, we consider only spherical shapes of the deformed bilayer
membrane, while introducing position independent (over the
membrane's surface) mean and gaussian curvatures:

\begin{eqnarray}
H=\frac{1}{r}\,;\;\;\; K=H^2=\displaystyle\frac{1}{r^2}
\label{curvs}
\end{eqnarray}

\noindent where $r$ is the radius of curvature. The radius of the
base of the curved membrane is fixed at $R\leq r$, in accord with
the fixed radius of the pipette, which sucks in the membrane, and
creates a pressure difference $P$ between inside and outside
surfaces of the membrane in the experimental setup \cite{4,5}.
According to experimental data \cite{4,5} the membrane under
consideration is rather thin:

\begin{equation}
l_s\sim 10^{-3} R\ll R\leq r \,.\label{thin}
\end{equation}

Choose direction "up" inside the pipette, perpendicular to the
plane of the initially flat membrane. Also, ascribe indices "1"
and "2" to the parameters of the lower and upper monolayer of the
membrane respectively. Then, the external surface areas $S_1$ and
$S_2$ of the lower (concave) and upper (convex) monolayers and the
numbers of the molecules in them are related by the following
equations:

\begin{eqnarray}
&&S_1=2\pi
(r-l_1)^2\left(1-\displaystyle\sqrt{1-\frac{R^2}{(r-l_1)^2}}\right)=
N_1a_1\;
;
\label{area1}\\
&&S_2=2\pi
(r+l_2)^2\left(1-\displaystyle\sqrt{1-\frac{R^2}{(r+l_2)^2}}\right)=
N_2a_2\;
;
\label{area2}\\
 && \pi R^2=N_0a_0\;, \mbox{at $r=\infty$}.  \label{area0}
\end{eqnarray}

\noindent Here the thicknesses $l_{i=1,2}$ of the i-th monolayer
are defined in Eqs. (\ref{curvs1}), (\ref{curvs2}). A flat,
undeformed membrane has a monolayer thickness $l_0=v/a_0$ under
zero external pressure gradient, where the area per molecule $a_0$
is determined from the condition of the minimum of the free energy
$F$ defined in Eq. (\ref{frent}):

\begin{eqnarray}
\displaystyle\frac{\delta F}{\delta a_0}=0\;,\;\;\;
\mbox{at:}\;\;\; a_1=a_2=a_0,\;\;\; H=K=0,\;\;\;
N_1=N_2=N_0.\label{a0}
\end{eqnarray}

Allowing for the experimental setup \cite{6}, it is reasonable to
suppose that the number of molecules in the upper monolayer $N_2$,
which "sticks" by its peripheral to the pipette walls, is fixed
during the experiment:

\begin{equation}
N_2=N_0. \label{n2}
\end{equation}

\noindent Hence, Eq. (\ref{n2}) together with Eqs. (\ref{area2}),
(\ref{curvs}), (\ref{curvs2}), and (\ref{lis}) relates curvature
radius $r$ to the area per molecule $a_2$:

\begin{equation}
a_2=a_2(H) \label{a2h}
\end{equation}

In order to find the equilibrium areas $a_1$ and $a_2$ one has to
solve different sets of self-consistency equations, depending on
the conservation/nonconservation of the number of molecules $N_1$
in the 1-st monolayer.

When $N_1$ is conserved we have:

\begin{equation}
N_1=N_0=N_2. \label{n11}
\end{equation}

\noindent  This equation, together with Eqs. (\ref{area1}),
 (\ref{curvs}), (\ref{curvs1}), and (\ref{lis}) relates $a_1$ to the
curvature $H$ similar to Eq. (\ref{a2h}):

\begin{equation}
a_1=a_1(H) \label{a1h}
\end{equation}

\noindent Hence, the only independent variables left are the
curvature radius $r$, and the pressure difference across e.g. 1-st
monolayer $P_1$. The equilibrium values of these parameters
minimize the free energy in the equilibrium state of the membrane
exposed to the total applied pressure difference $P$:

\begin{eqnarray}
\displaystyle\frac{\delta F_i}{\delta
a_i}=T_i=\frac{P_ir}{2};\;\;P_1+P_2=P;\;\;\;P_1\approx
P_2=\frac{P}{2}. \label{tis}
\end{eqnarray}

\noindent Here the tensions $T_i$ (force per unit length of the
peripheral) acting in the monolayers are related to the
corresponding pressure differences $P_i$ via the Laplace's law
(\ref{laplace}). A distribution of the pressure differences $P_i$,
acting across each monolayer, generally should be determined
self-consistently by a minimization of the membrane's free energy
$F$, but for a thin membrane (see (\ref{thin})) we choose them
approximately equal.

In the case, when $N_1$ is not fixed and the 1-st monolayer is
free to absorb extra lipid molecules from the patch/reservoir
through its peripheral edge, we substitute equilibrium conditions
(\ref{tis}) by the different ones:

\begin{eqnarray}
&&\displaystyle\frac{\delta \tilde{F}}{\delta a_1}=T_1=0\;;
\label{ti1}\\
&&\displaystyle\frac{\delta \tilde{F}}{\delta
a_2}=T_2=\frac{Pr}{2}\;;
\label{ti2}\\
&&\tilde{F}=F+F_{\mbox{patch}}\;, \;\;\;\;
N_1+N_2+N_{\mbox{patch}}=const\label{totalf}
\end{eqnarray}

\noindent where for simplicity the patch free energy
$F_{\mbox{patch}}$ is assumed to be equal to the energy of a flat
piece of the bilayer membrane under zero pressure and zero
tension, which contains $N_{\mbox{patch}}$ lipid molecules in
total:

\begin{eqnarray}
F_{\mbox{patch}}=\sum_{i=1,2}\displaystyle\frac{1}{2}
{N_{\mbox{patch}}}f_0;\;\;
f_0=f_1=f_2\;\; \mbox{at:}\;\;\; a_1=a_2=a_0,\;\;\; H=K=0.
\label{frepatch}
\end{eqnarray}

\noindent The free energies $f_i$ are defined in (\ref{fren}).
Finally, we solve equations (\ref{tis}) and (\ref{ti1}),
(\ref{ti2}). As a result, we determine the final state with the
lowest energy by a direct comparison of the values of the free
energy $\tilde{F}$ in the two different cases of the
"patch-disconnected" and "patch-connected" membrane.

\section{Solutions}

It is convenient for calculational purposes to rewrite expressions
obtained in the previous Section using dimensionless parameters
instead of $a_i$ and $H$. Because the expressions for the free
energy (\ref{frent}), (\ref{frepatch}) will be studied mainly
numerically, we shall incorporate tensions $T_i$ into the free
energy and look for the minima of the latter on the proper
manyfold of the independent variables, rather than solve directly
equations (\ref{tis}) - (\ref{ti2}) of the Euler-Lagrange type.
For this purpose we pass from the free energies $f_i$ to the new
ones $\tilde{f}_i$, which are Helmholtz free energies per lipid
molecule in the i-th monolayer under an external pressure
difference $P_i$:

\begin{eqnarray}
&&\tilde{f}_i=f_i(z_i(h),h)-\displaystyle
\int^{z_i(h)}_{z_0}\frac{P_iv}{2h_i(z')}dz'\;;
\label{fretildei}\\
&&f_1(z_1,h)\equiv
\epsilon_sz_1+\epsilon_t\left\{\displaystyle
\frac{h^2}{3}(7z_1^{-4}-4z_1^{-3})+
2h(z_1^{-2}-z_1^{-3})+1+z_1^{-2}-2z_1^{-1}\right\};
\label{fredimless1}\\
&&f_0=f_1(z_0,h=0)=\epsilon_sz_0+
\epsilon_t\{1+z_0^{-2}-2z_0^{-1}\}; \label{fredimless2}\\
&&f_2(z_2,h)=f_1(z_2,-h)\label{fredimless0}.
\end{eqnarray}

\noindent Here dimensionless parameters $z_i$ and $h$,
as well as
the other ones, are defined as follows:

\begin{eqnarray}
&&\displaystyle z_i=\frac{a_il_s}{v};\;\;h=\frac{l_s}{r}\equiv
Hl_s;\;\;h_0=\frac{l_s}{R};\;\;\epsilon_s=\frac{\gamma
v}{l_s};\;\;\epsilon_t=\frac{k_sl_s^2}{2}.\label{dimless}
\end{eqnarray}

\noindent A dependence $h_i(z)$ on the right-hand-side of Eq.
(\ref{fretildei}) is just the inverse of either Eq. (\ref{a1h}) or
Eq. (\ref{a2h}) written in the dimensionless units
(\ref{dimless}). It is useful for the comprehension of the rest of
the paper to mention here orders of magnitude of the main
parameters. As it follows from (\ref{thin}): $h\sim 10^{-3}\ll 1$.
Next, it follows from the definition (\ref{lis}) and Eqs.
(\ref{curvs1}), (\ref{curvs2}) that $v\approx l_s a_i $, which
leads to an estimate: $z_i\sim 1$. The entropic nature of the
$f_{ti}$ term in Eq. (\ref{fren}) is reflected in the temperature
dependent coefficient $k_s$, which for the known phospholipid
bilayers is of the order \cite{7}: $k_sl_s^2\sim 1\div 10 kT$,
where $k$ is the Boltzman's constant and $T$ is the absolute
temperature. The scale of the surface term $f_{si}$ in Eq.
(\ref{fren}) is characterized by the coefficient $\gamma$, which
at room temperature is of the order \cite{7}: $\gamma\sim 0.1
kT/\AA^2$. Gathering all the estimates, and also taking into
account that the typical value of an area per molecule is
\cite{7}: $a_i\sim 60\AA^2$, we obtain the following list of the
estimates (taken for the room temperature $T\sim 300K$):

\begin{eqnarray}
&&\displaystyle z_i\sim 1;\;\;h\sim 10^{-3};\;\;\epsilon_s\sim
10^{-13}\mbox{erg};\;\; \epsilon_t\sim 1\div 10
\epsilon_s.\label{dimvalues}
\end{eqnarray}

 \noindent An additional important dimensionless parameter relates
 characteristic experimental value of the pressure difference $P
 \sim 50$mm Hg \cite{4,5} with the "microscopical bending energy"
 $\epsilon_t$:

\begin{eqnarray}
\displaystyle\frac{P v}{2\epsilon_t}\sim 4\cdot
10^{-4}\;\;\;\mbox{at:}\;\;\epsilon_t\sim 10^{-13}\mbox{erg};\;\;
v\sim 1.2\cdot 10^{3}\AA^{3} , \label{dimp}
\end{eqnarray}

\noindent where we use an estimate $l_s\sim 20\AA$ for the typical
length of a phospholipid molecule in a bilayer membrane \cite{7}.

\vspace{3mm} Now, coming to a description of the results, we first
solve equation for $z_0$:

\begin{eqnarray}
\displaystyle\frac{\partial{f_0}}{\partial{z_0}}=\epsilon_s+
2\epsilon_t\{z_0^{-2}-z_0^{-3}\}=0
 \label{z0}
\end{eqnarray}

\noindent or in the Cardano's form:

\begin{eqnarray}
&&z_0^3+pz_0+q=0; \;\;\mbox{where:}\;\; \displaystyle
p=-q=\frac{2\epsilon_t}{\epsilon_s}\label{cardano}\\
&&\mbox{and:}\;\;
\displaystyle\left(\frac{q}{2}\right)^2+
\left(\frac{p}{3}\right)^3\equiv
\left(\frac{\epsilon_t}{\epsilon_s}\right)^2+
\left(\frac{2\epsilon_t}{3\epsilon_s}\right)^3>0.
\label{inequal}
\end{eqnarray}

\noindent The inequality in (\ref{inequal}) proves that there
is a
unique real root of the cubic equation (\ref{cardano}). In the
limit $p\gg1$ the real root of Eq. (\ref{cardano}), first order
accurate in $p^{-1}$, is:

\begin{eqnarray}
z_0\approx
1-\frac{1}{p}=1-\displaystyle\frac{\epsilon_s}{2\epsilon_t}.
\label{cardanosol}
\end{eqnarray}

\subsection{Patch-disconnected case}

In the patch-disconnected case, or during a short enough time $t$:
$\Delta t\leq t\ll\tau$, after an application of a pressure
difference $P$ across the membrane's plane, the numbers of the
lipid molecules in the both monolayers of the bilayer membrane
remain equal: $N_1=N_2=N_0$. Hence, the areas per molecule
$a_{i=1,2}$ also remain equal to each other, while simultaneously
undergoing an increment with respect to the initial undeformed
value $a_0=vz_0/l_s$. Neglecting small corrections $\propto
l_i/r\sim 10^{-3}$ in Eqs. (\ref{area1}), (\ref{area2}) one finds:

\begin{eqnarray}
z_i(h)\approx z_0
\frac{2h_0^2}{h^2}\left(1-\displaystyle
\sqrt{1-\frac{h^2}{h_0^2}}\right);\;\;
i=1,2. \label{ais}
\end{eqnarray}

\noindent Thus, $a_i$ ranges from $a_0$ to $2a_0$, or
equivalently: $z_0\leq z_i(h)\leq 2z_0$, on the interval of the
membrane's curvatures: $0\leq 1/r\leq 1/R $, or in dimensionless
units: $0\leq h\leq h_0$. The inverse of Eq. (\ref{ais}) yields:

\begin{eqnarray}
h(z_i)\approx
\frac{2h_0z_0}{z_i}\displaystyle\sqrt{\frac{z_i}{z_0}-1};
\;\;i=1,2. \label{hais}
\end{eqnarray}

\noindent Now, we substitute (\ref{hais}) into the integral on
the
right-hand-side of (\ref{fretildei}) and find after an
integration:

\begin{eqnarray}
&&\tilde{f}_i=f_i(z_i(h),h)-\displaystyle
\frac{P_ivz_0}{2h_0}\sqrt{\frac{z_i(h)}{z_0}-1}
\left[1+\frac{1}{3}\left(\frac{z_i(h)}{z_0}-1\right)\right].
\label{fretildeI}
\end{eqnarray}

\noindent Substituting $P_i=P/2$ in (\ref{fretildeI}), and then
substituting $\tilde{f}_i$ from Eq. (\ref{fretildeI}) into Eq.
(\ref{frent}) instead of $f_i$ we find the $F(h)/N_0$ dependence
(in the units of $\epsilon_t$) plotted with the dashed curve in
Fig. 2. A position of the minimum, defined as the point at which
$\partial F/\partial h$ changes sign, is indicated with the
vertical arrow. It is apparent from the Fig. 2 that the minimum is
rather shallow and happens at the curvature radius $r$ comparable
with the radius $R$ of the pipette (e.g. of the membrane's base).
This indicates a substantial deviation of the membrane from the
planar shape and is in accord with the experiments demonstrating
nearly spherical shape of the membrane at this pressure
\cite{4,5}. Thus, it is not surprising that, unlike in the case of
a weakly bent thin plate \cite{LL}, now both sides of the
membrane's surface are stretched. Hence, an important property of
this deformed state is the presence in both monolayers of nearly
equal tensions $T_i$ stretching the membrane and causing opening
of the protein mechanosensitive channels \cite{4,5}: $T_1\approx
T_2=Pr/4$. Let us consider another situation, where the symmetry
of the tension distribution is broken.

\subsection{Patch-connected case}
Now assume that enough time had elapsed after the beginning of the
pressure pulse for a transfer of a substantial amount of the lipid
molecules from the patch to the "sucked in the pipette" part of
the membrane (see Fig. 1): $t\geq \tau$. Hence, we consider the
number of molecules $N_1$ in the patch-connected monolayer as an
additional variational parameter. In this case the system of
equations (\ref{ti1}), (\ref{ti2}) applies. The first equation
written in the dimensionless variables and allowing for the
general expression (\ref{fredimless1}) yields:

\begin{eqnarray}
\epsilon_s+\epsilon_t\displaystyle\left\{\displaystyle
\frac{h^2}{3}
\left(-\frac{28}{z_{1}^{5}}+\frac{12}{z_{1}^{4}}\right)
+2h\left(\frac{3}{z_{1}^{4}}
-\frac{2}{z_{1}^{3}}\right)+2\left(\frac{1}{z_{1}^{2}}
-\frac{1}{z_{1}^{3}}\right)
\right\}=0 \label{varfre1}.
\end{eqnarray}

\noindent This equation defines now a $z_1(h)$ for the
tension-free monolayer with a finite curvature $h\neq 0$, and
thus, substitutes equation (\ref{z0}), which is valid for the
tension-free, but flat case ($h=0$). Simultaneously, equation
(\ref{ais}) remains valid for the 2-nd monolayer with conserved
number of lipids $N_2=N_0$, and provides dependence $z_2(h)$:

\begin{eqnarray}
z_2(h)\approx z_0
\frac{2h_0^2}{h^2}\left(1-\displaystyle
\sqrt{1-\frac{h^2}{h_0^2}}\right).
\label{ais2}
\end{eqnarray}

In the Fig. 3 we present our numerical solutions of the Eqs.
(\ref{z0}) (dash-dotted curve) , (\ref{varfre1}) (solid curves),
as well as a dependence given by Eq. (\ref{ais2}) (dotted curve).
The insert in Fig. 3 demonstrates that actually $z_1(h)$ remains
nearly equal to $z_0$ at all curvatures $h$. This result is not
surprising. Namely, equations (\ref{z0}) and (\ref{varfre1})
differ only by $\propto h^2\ll 1$ and $\propto h\ll 1$ terms. This
means, that the tension-free, relaxed 1-st monolayer, in order to
reach its final equilibrium state, absorbs as much molecules from
the patch as necessary to compensate for an increase of its area
at a finite curvature. In this way the area per molecule remains
nearly constant and equal to $\approx z_0$ (see insert in Fig. 3).
Then, as in the flat case, the equilibrium value $z_1$ is again
determined by the balance of the surface tension and entropic
repulsion of the lipid tails. Using dimensionless area $z_1(h)$
determined from (\ref{varfre1}), we find the (variable) number of
the lipid molecules $N_1(h)$, normalized by $N_0$, as a function
of curvature $h$, see Fig. 4. This result was obtained using the
following formula:

\begin{eqnarray}
 \displaystyle\frac{N_1}{N_0}= \displaystyle 2\frac{z_0}{z_1(h)}
 \left(\frac{h_0}{h}\right)^{2}\left(1-\sqrt{1-
 \left(\frac{h}{h_0}\right)^2}\right),
\label{n1}
\end{eqnarray}

\noindent where equation (\ref{area1}) and definitions
(\ref{dimless}) where used.

Now, in order to find the equilibrium value of the curvature $h$
(indicated with the arrow in Figs. (3),(4)) at a given pressure
difference $P$, we calculate the total free energy $\tilde{F}$
from Eq. (\ref{totalf}) normalized by $N_0$ using the following
expression:

\begin{eqnarray}
&&\frac{\tilde{F}}{N_0}=f_1(z_1(h),h)\frac{N_1}{N_0}+f_2(z_2(h),h)-
\displaystyle\frac{Pvz_0}{2h_0}\sqrt{\frac{z_2(h)}{z_0}-1}
\left[1+\frac{1}{3}\left(\frac{z_i(h)}{z_0}-1\right)\right]-
\nonumber \\
&& \displaystyle f_0\left(\frac{N_1}{N_0}-1\right). \label{frepat}
\end{eqnarray}

\noindent The physical meaning of the factor $N_1/N_0$ in front of
$f_1$ in combination with the last subtracted term $\propto f_0$
in (\ref{frepat}) is simple. Namely, the total number of molecules
in the system membrane$+$patch is conserved. Therefore, an
increase of the number of molecules in the first monolayer, with
the free energy per molecule $f_1$, occurs at expense of the
decrease of the number of molecules in the patch, with the energy
per molecule $f_0$. Calculation of the free energy
${\tilde{F}}/{N_0}$, normalized by $\epsilon_t$, provides its
dependence on $h$ expressed with the solid curve in Fig. 2. The
arrow indicates position of the minimum of the free energy
(obtained from the calculation of the zero root of the first
derivative of the free energy with respect to the curvature).

We have also calculated the dependencies of the dimensionless
equilibrium curvature $h=R/r$ on the applied pressure difference
$P$ for the symmetric strain (dashed line): $T_1=T_2$, and
asymmetric strain (solid line): $T_1=0, T_2\neq 0$, equilibrium
states, see Fig. 5. As it follows from the figure, the power-law
dependencies prove to be the same: $1/r\sim const\cdot
P^{\alpha},\; \alpha \approx 1/3$, but with the constant
pre-factor slightly smaller in the symmetric case.

The influence of the effective surface tension energy
$f_{si}=\gamma a_i$ in Eq. (\ref{fren}), characterized by the
energy $\epsilon_s$ (\ref{dimless}), is demonstrated in Fig. 6.
Here, it is obvious, that an increase of the surface energy
coefficient at a fixed entropic repulsion $\epsilon_t$, causes a
greater bending rigidity of the membrane and as a result, a
decrease of the equilibrium curvature caused by an applied
external pressure difference.

\section{Conclusions and experimental proposals}

A direct comparison of the two curves in Fig. 2 leads to the main
conclusion of this work, i.e. the free energy of the relaxed,
partially tension-free asymmetric equilibrium state: $T_1=0,
T_2=Pr/2$ is {\it{lower}} than the free energy of the non-relaxed
symmetric equilibrium state: $T_1\approx T_2=Pr/4$. Thus, the
membrane would change its state until the equilibrium state with
the lowest free energy would be reached. The number of the lipids
in the 1-st monolayer $N_1$ in the relaxed state is greater than
the initial one $N_0$ by a macroscopical amount ($\sim 15\% N_0$,
see Fig. 4). Hence, a process of reaching the relaxed state would
take time necessary for diffusion of the molecules from the patch
into the 1-st monolayer through its peripheral edge (see Fig 1).
This time, $\tau$, as evaluated in Section II, Eq.
(\ref{diffusion}), is of order of {\it{seconds}}, and thus is in
correspondence with the characteristic time of the collapse of the
ionic conductance through the protein mechanosensitive channels
piercing the E-coli's membrane \cite{5}. Hence, our theory
confirms the idea \cite{6}, that an inter-layer slide in a
membrane, soon after the beginning of the pressure pulse, may
cause a relaxation of the lateral stress. This stress had
initially opened the protein mechanosensitive channels and thus
had given rise to the ionic conductance. Closing back of the
channels within seconds after the beginning of the pulse would
cause simultaneous decrease of the ionic conductance as seen
experimentally \cite{5}. A separate consideration of the two
equilibrium states discussed above, is justified by the fact that
the characteristic stress relaxation time: $\tau\sim 1 sec$
differs by 5$\div$6 orders of magnitude from the characteristic
time $\Delta t$ necessary for a propagation of the bending
deformations over a membrane with the linear dimensions $R\sim
1\mu m$:

\begin{eqnarray}
 \displaystyle \Delta t\sim R^{2}
 \left(\frac{\rho l_0}{\epsilon_t}\right)^{1/2}
 \sim 10^{-8}cm^{2}\left(\frac{1g/cm^{3}
 \cdot 10^{-7}cm}{10^{-13}erg}\right)^{1/2}
 \sim 10^{-5}sec\;,
\label{deltat}
\end{eqnarray}

\noindent where we used a well known formula for the frequency of
the bending waves in a thin plate \cite{LL}:
 \[\omega=q^2\left(\frac{k_B}{\rho l_0}\right)^{1/2},
 \]

\noindent where $k_B$ is a bending modulus, $\rho$ is a mass
density, $l_0$ is a thickness, and $q$ is a wave-vector. We also
allowed for the role of a bending coefficient $k_B$ played by the
energy $\epsilon_t$, which follows from direct comparison of the
free energy expressions (\ref{ftis1}), (\ref{ftis2}) and the
definition of $\epsilon_t$ given in (\ref{dimless}).

\vspace{3mm} Based on our results, shown in Figs. 2, 4, we propose
here the following possible experimental verifications of the
theory. First, it is obvious from the Fig. 2 that the equilibrium
curvature is {\it greater} in the relaxed state than in the
non-relaxed one, changing correspondingly from $1/r=0.525 R^{-1}$
in the "intermediate" equilibrium state to $0.675 R^{-1}$ in the
relaxed state. Hence, we propose to measure this change
experimentally as an accompanying process of the decrease of the
ionic conductance. This change of curvature signifies a simple
fact. Shortly after the application of the pressure difference $P$
it is balanced by the two strained monolayers of a membrane. After
enough time has elapsed, a relaxation of the strain in the 1-st
monolayer occurs, and hence, only the strain in the 2-nd monolayer
compensates the applied pressure, thus the curvature has to
change.

Second, from Fig. 4 it follows that, as a result of the relaxation
of the strain in the 1-st monolayer, this monolayer acquires extra
lipid molecules from the patch, hence the number of molecules in
the pipette increases by: ($\Delta N_1\approx 15\% N_0$). As this
change $\Delta N_1$ is macroscopic, a corresponding increase of
the weight of the sucked in membrane {\it{during}}, the decrease
of the ionic conductance should be, in principle, measurable.

Finally, we mention some possible future improvements of the
theory. These could be made by lifting of our simplifying
restrictions of the homogeneity of the lateral stresses (tensions)
across the thickness of each monolayer, as well as the
approximation of a constant area per molecule over each external
surface of the membrane. As a consequence, in a more complete
scheme of derivations one would like not to restrict himself to
the spherical symmetry of the membrane's shape. Nevertheless, we
believe that these improvements can not overturn the main physical
concept considered in this paper of a diffusion mechanism of a
stress relaxation in a bilayer membrane with an inter-layer slide.

\section*{ACKNOWLEDGEMENTS}

The authors are grateful to S.I. Sukharev for the introduction in
the problem and for numerous enlightening discussions during the
work. Useful discussions with Jan Zaanen and comments by R.F.
Bruinsma and Th. Schmidt are highly acknowledged. S.I.M. is
grateful to R.A. Ferrell for the warm hospitality during stay in
Maryland and to V.M. Yakovenko for useful comments.

\newpage
\section{Figure Captions}

Fig. 1. Sketch of the experimental setup \cite{4}. The numbering
1,2 of the monolayers within a bilayer membrane is indicated.
\vspace{5mm}

Fig. 2. Free energy $F/{(\epsilon_t N_0)}$ normalized by
$\epsilon_t$ and number of lipid molecules $N_0$ {\it{versus}}
dimensionless curvature $h$. Solid curve corresponds to the
asymmetric strain state with $N_2=N_0<N_1$. Dashed curve
corresponds to the symmetric strain case with conserved numbers of
lipids in the monolayers: $N_1=N_2=N_0$. Arrows indicate positions
of the minima of the free energy.

\vspace{5mm} Fig. 3. Dimensionless areas per molecule $z_i$ in the
i-th monolayer of a bilayer membrane as functions of the
membrane's curvature $h$ in the asymmetric strain state
$T_1=0,T_2=Pr/2$ (see text). Arrow indicates equilibrium value of
the curvature. $z_0$ is the equilibrium value under zero external
pressure difference.

\vspace{5mm} Fig. 4. Calculated number of the lipid molecules
$N_1$ in the first monolayer as a function of curvature $h$ in the
asymmetric strain state $T_1=0,T_2=Pr/2$. Arrow indicates
equilibrium value of the curvature at $Pv/(2\epsilon_t)=0.0004$.

\vspace{5mm} Fig. 5. Calculated dependencies of the dimensionless
equilibrium curvature $h=R/r$ on the applied pressure difference
$P$ for the symmetric strain (dashed line) and asymmetric strain
(solid line) equilibrium states in the logarithmic scales.

\vspace{5mm} Fig. 6. Free energy $F/{(\epsilon_t N_0)}$ normalized
by $\epsilon_t$ and number of lipid molecules $N_0$ {\it{versus}}
dimensionless curvature $h$. All notations are as in Fig. 2. The
main panel corresponds to the ratio $\epsilon_t/\epsilon_s=50$,
which is 10 times greater than for the Fig. 2. An insert
calculated for the ratio $\epsilon_t/\epsilon_s=0.05$. Arrows
indicate equilibrium values of the curvature.

 %%%%%%%%%%%%%%%%%%%%%%%%%%%%%%%%%%%%%%%%%%%%%%%%%%%%%%%%%%%%%%%%
\newpage
\begin{figure}
 \vbox to 4.0cm {\vss\hbox to -5.0cm
 {\hss\
       {\includegraphics{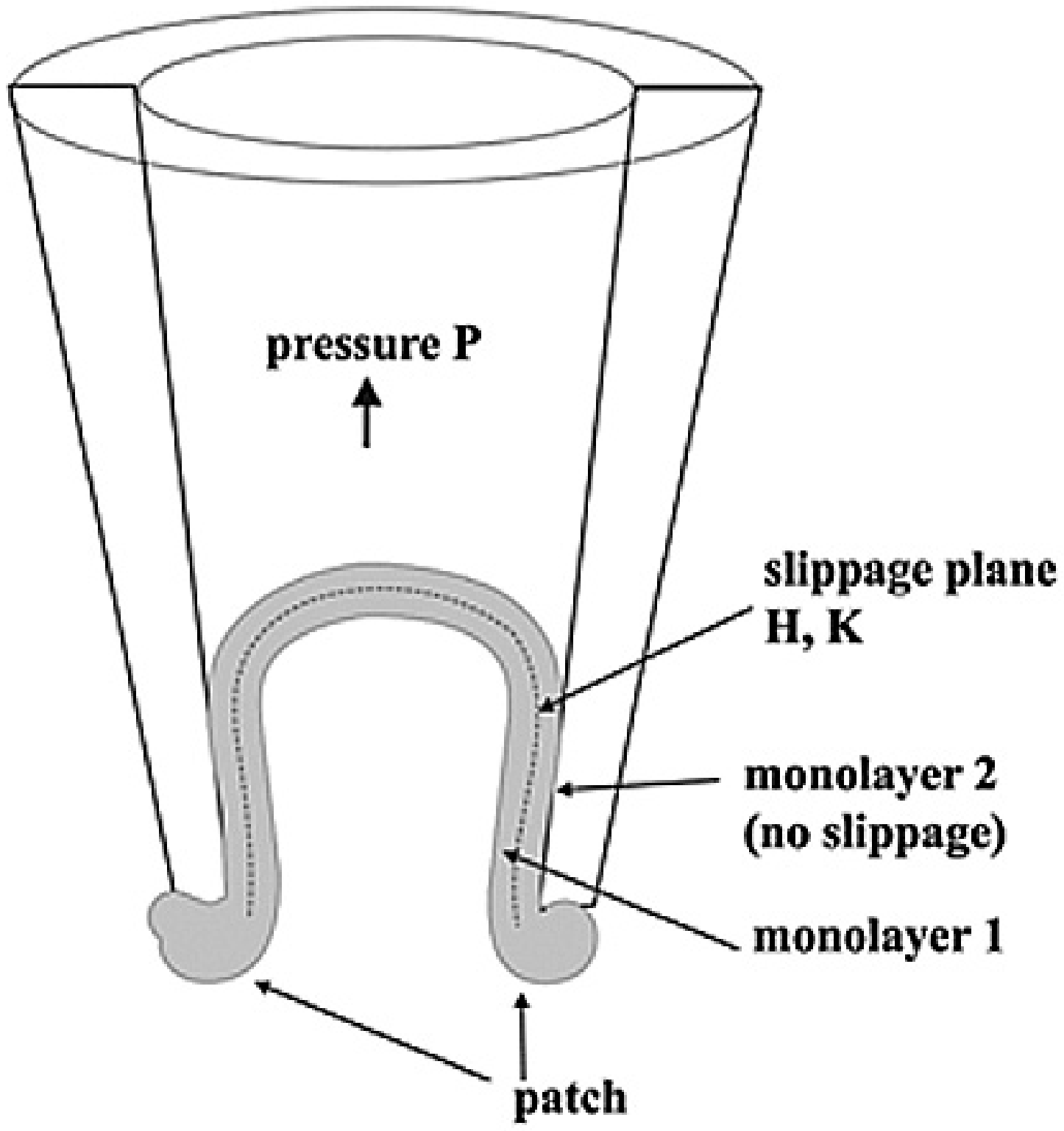}
       }
  \hss}
 }
\vspace{17cm} \caption{}  \label{patch}
\end{figure}

\newpage

\begin{figure}
 \vbox to 4.0cm {\vss\hbox to -5.0cm
 {\hss\
       {\includegraphics{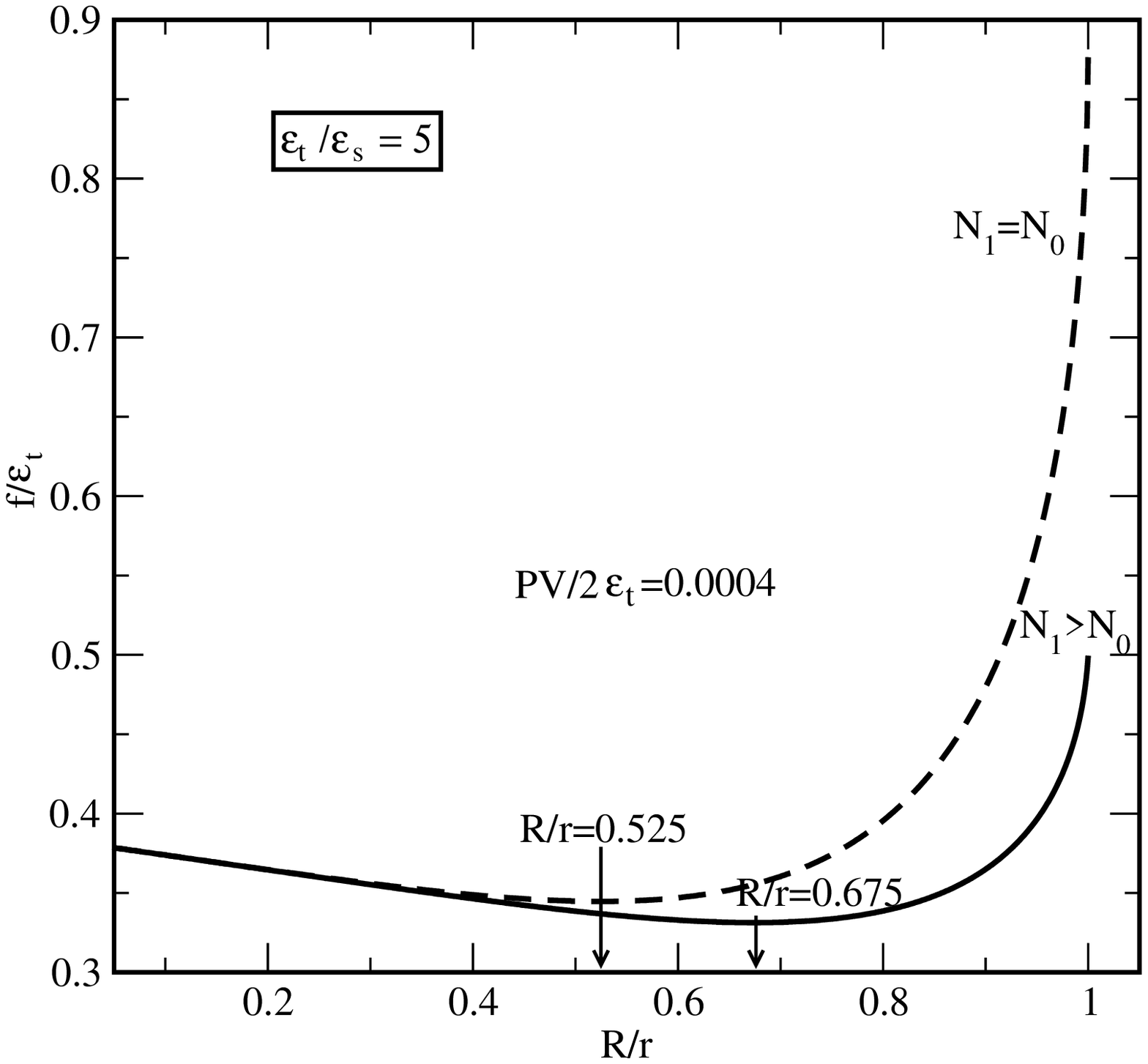}
       }
  \hss}
 }
\vspace{18.5cm}\caption{} \label{frens}
\end{figure}
\newpage

\begin{figure}
 \vbox to 4.0cm {\vss\hbox to -5.0cm
 {\hss\
       {\includegraphics{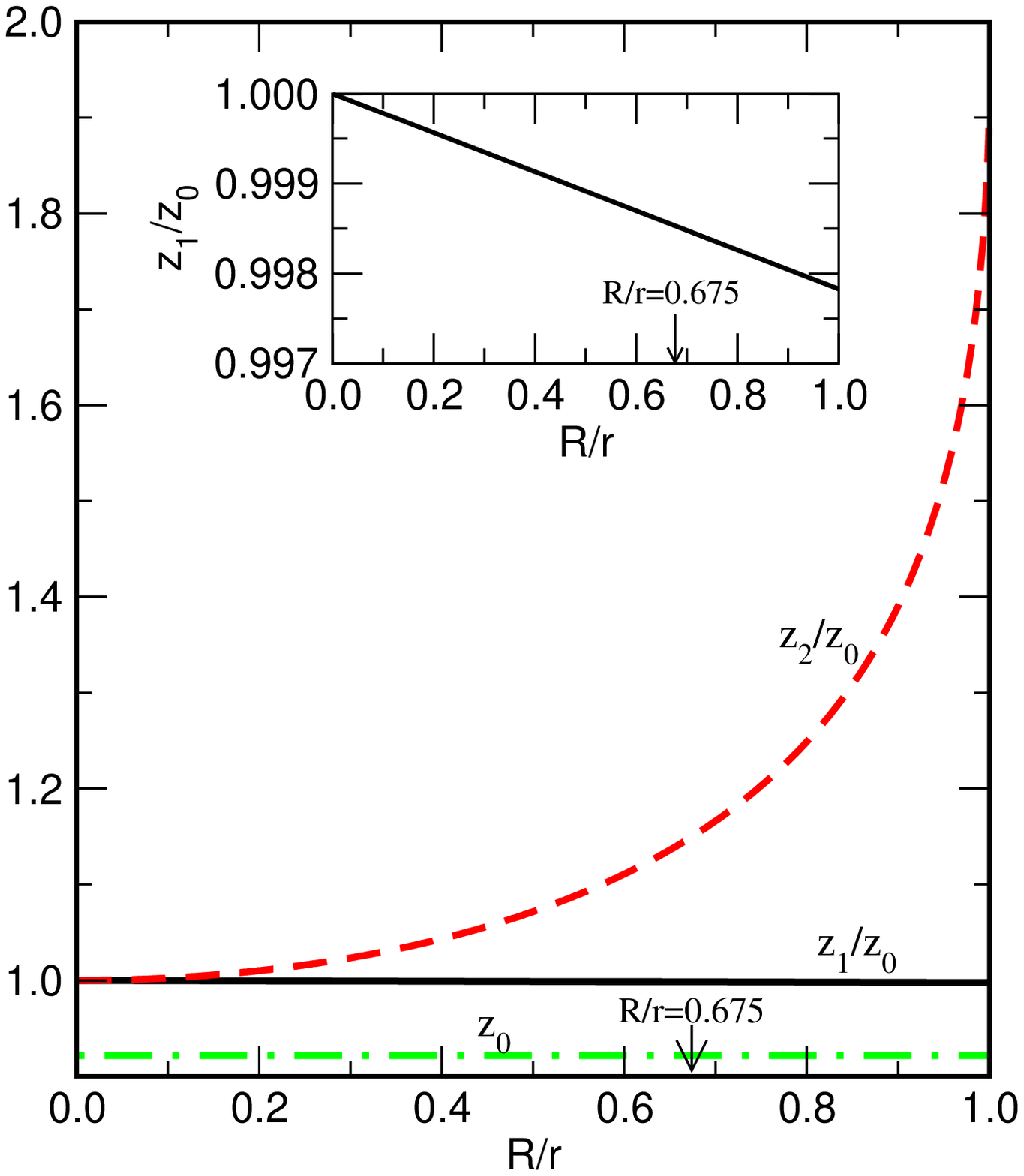}
       }
  \hss}
 }
\vspace{17.5cm}\caption{} \label{zis}
\end{figure}
\newpage

\begin{figure}
 \vbox to 4.0cm {\vss\hbox to -5.0cm
 {\hss\
       {\includegraphics{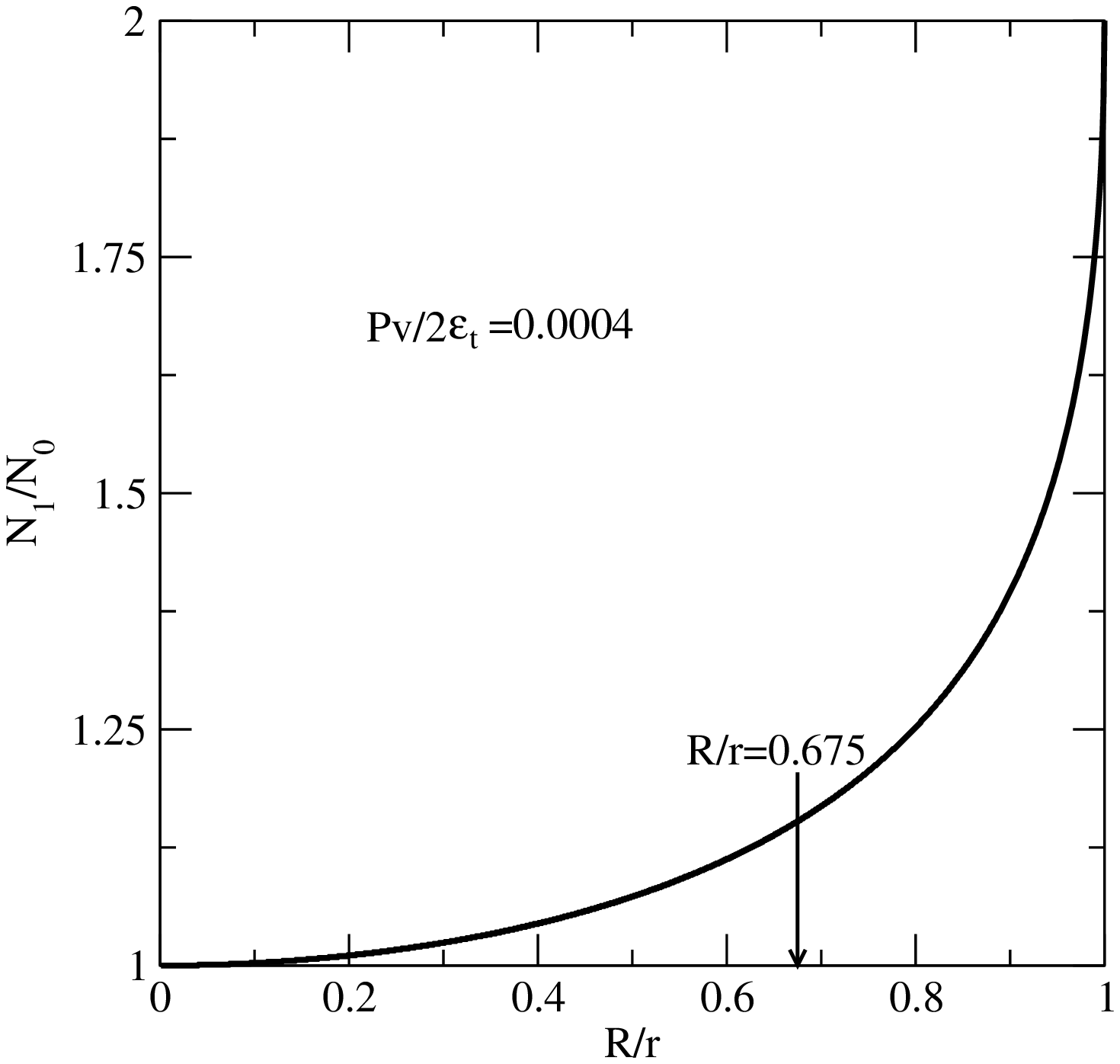}
       }
  \hss}
 }
\vspace{17.5cm}\caption{} \label{layn1}
\end{figure}

\newpage

\begin{figure}
 \vbox to 4.0cm {\vss\hbox to -5.0cm
 {\hss\
       {\includegraphics{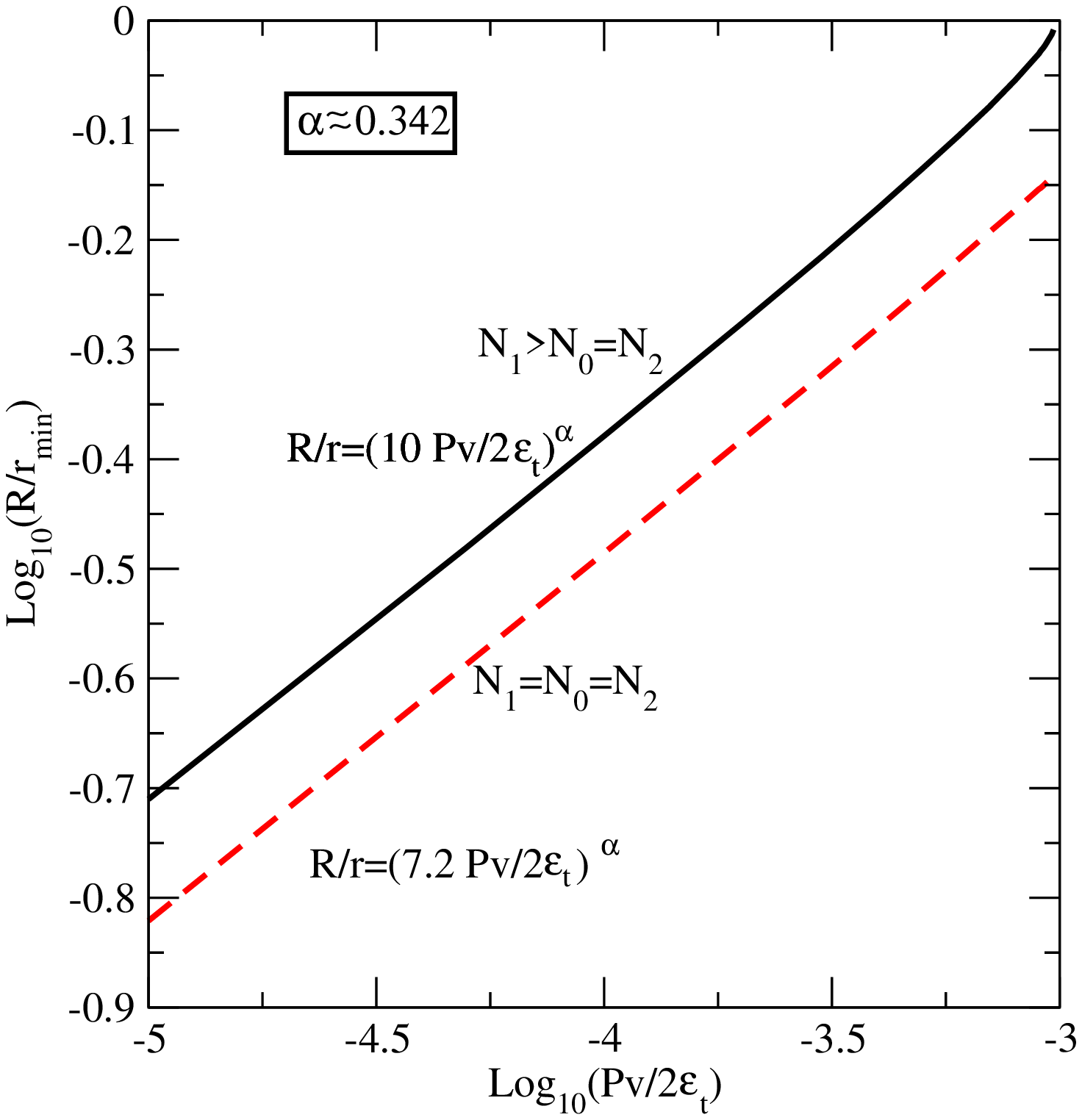}
       }
  \hss}
 }
\vspace{17.5cm}\caption{} \label{loghlogp}
\end{figure}

\newpage

\begin{figure}
 \vbox to 4.0cm {\vss\hbox to -5.0cm
 {\hss\
       {\includegraphics{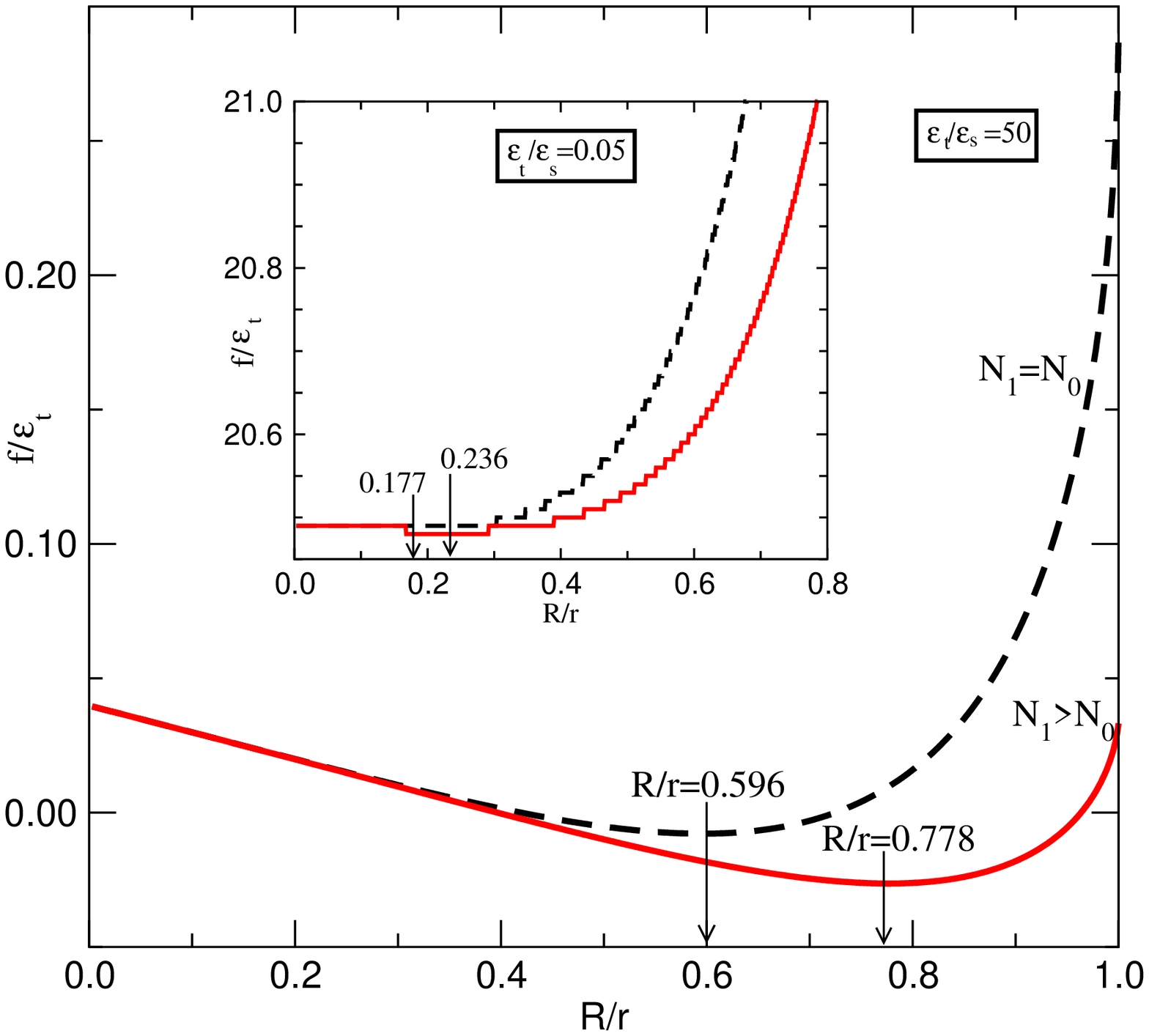}
       }
  \hss}
 }
\vspace{17.5cm}\caption{} \label{frios1}
\end{figure}

\end{document}